\documentstyle[sprocl]{article}
\def\be{\begin{equation}} 
\def\ee{\end{equation}}
\def\bea{\begin{eqnarray}}
\def\eea{\end{eqnarray}}

\begin{document}
\title{QCD - looking forward}

\author{L. Frankfurt}

\address{School of Physics and Astronomy,
 Tel Aviv University,\\ Tel Aviv 69978, Israel}

\author{M.Strikman}

\address{Department of Physics, Pennsylvania State University, \\
University Park, PA 16802, USA}
\maketitle

\centerline{Abstract}

We outline  theoretical ideas on the soft and hard dynamics of strong  
high-energy  interactions and discuss promising directions for 
future high-energy experimental investigations including the ones 
which would allow one  to reveal the three-dimensional structure of  QCD
bound states, investigate the onset of regime of large parton densities,
and  observe the violation of the DGLAP evolution equation.
We emphasize that many  qualitatively new phenomena should be
present for the forward kinematics both in electron-nucleon(nucleus)
collisions at HERA and in  $pp, pA$ collisions at LHC.

\section{Introduction}

Our current understanding of Nature - the Standard Model of
Strong and  Electroweak  interactions - is formulated in terms of 
interactions of Abelian and non-Abelian
gauge fields. All these Lagrangians lead to intricate phenomena in 
the situations where the coupling constant becomes large. One can even 
question whether these models are selfconsistent in the nonperturbative
regime.  
The most pressing theoretical  problems are the dynamics relevant for
the confinement of quarks and gluons,  the phenomenon of spontaneously
broken continuous  symmetries, the role of nearly massless quarks in 
QCD confinement, the possible existence of new forms of stable and 
metastable  matter, kinetics of phase transitions, as well as the
role of nonlinear chaotic phenomena in the large 
parton density regime. Obviously, experimental
investigations of these problems in a non-Abelian gauge theory in 
perturbative and nonperturbative regimes in a Lab is feasible for 
QCD only. Hence the  importance of QCD research for  physics as a whole.
Note also that the current analyses of the LEP data reviewed in the 
summary talk by J.Ellis ~\cite{Ellis} make it quite probable that the 
Higgs meson is light enough to be 
discovered
before the 
LHC, and supersymmetry particles are too heavy to be discovered at the LHC.
In this case, QCD may provide the only avenue for discoveries
of new phenomena at the  LHC.

Recent HERA data observed a  fast increase of parton distributions 
in DIS at small $x$  predicted by 
perturbative QCD \cite{smallxincrease}.  There are about  19 gluons 
in a proton at HERA energies and the gluon density of a proton
increases with energy. Furthermore, HERA started investigating
exclusive hard diffraction processes which are strikingly 
different from  soft QCD diffraction, but are in line with the 
predictions based on the generalization of the QCD factorization 
theorem to  exclusive DIS processes. Related experimental 
studies at FNAL of the diffractive dissociation of a pion into two 
high $p_t$ jets ~\cite{Ashery}, of $J/\psi$ photo production off
nuclear targets ~\cite{sokoloff} with a coherent nuclear recoil 
indicate the  existence of a new QCD phenomenon: transparency of nuclear 
matter to  the propagation of a fast spatially small wave package of quarks 
and gluons, cf. discussion and references in ~\cite{FMS}.   
Thus,  one is tempted to look for  ways to experimentally  reach new
hard-soft QCD regimes of large parton densities which may be
manifested in a variety of Color Coherent Phenomena. The aim of 
this talk is to outline new and nontrivial QCD 
phenomena which can be discovered under  the extreme 
conditions achievable in the kinematics of 
small $x$ processes in the  LHC energy range and at HERA
in  both  $ep$ and $eA$ modes.

~

{\it Challenging Problems}

$\bullet $ ~~Does  a spatially small color singlet  object 
interact 
with a hadron target, $T$, at high energies with a  small cross
section, or  can this cross
section  reach a maximum strength allowed by the geometric
($s$-channel) constraint of $\approx \pi r^2_T$ as hinted by 
the  naive extrapolation of DGLAP calculations into the small $x$ region. 
What is the origin of the interaction strength fluctuations for large
interaction strengths. How do two small objects interact at small
$x$.
 \footnote{The last problem is often referred to as the BFKL model 
of the Pomeron. This subject is discussed at  length in the talk of
A.Mueller ~\cite{Muellerdis98} and hence we will skip it in this talk.}

$\bullet$ ~~
What dynamics governs  the states at high parton densities produced 
in hard and in soft collisions. How does one  produce and investigate
various new forms of hadron matter using well understood QCD phenomena.

$\bullet$ ~~How can one extend 
the well understood methods of measuring the  single parton densities 
to study  multiparton correlations including
correlations between longitudinal and transverse degrees of freedom.
The ultimate goal would be to extract  
the three-dimensional image of the nucleon from appropriate data.

\section{Perturbative QCD in the interactions of small dipoles with hadrons}
\subsection{Cross section of dipole-hadron interaction}
A number of hard diffractive phenomena 
can be expressed through the interaction cross section of a small color dipole
with a hadron target using the  QCD factorization theorem.
The LO expression for this cross section is  ~\cite{Blattel93}:
\be
\sigma_{dipole ~ T}(x,b)=\pi^2 {F^2\over 4} b^2 
\alpha_s(b_o^2/b^2) x G_T(x,b_o^2/b^2)   
\label{plc}
\ee
In the case of a $q \bar q$ pair, $\vec{b}$ is the relative
impact parameter between quark and antiquark:
$(\vec{b} \cdot \vec{P})=0$; 
$x={\lambda\over sb^2}$ and $F^2$ is the  Casimir 
operator of $SU(3)_c$.

~

{\it Implications of eq.\ref{plc}:}

~

$\bullet$ ~~ 
Color screening property of eq.\ref{plc} - $\sigma(b)_{|b\to 0} \to 0$
in combination with the phenomenon of Lorentz slowing down
of the space-time evolution for a fast particle
interaction allows one  to justify 
the applicability of the factorization theorem
to  hard diffractive phenomena initiated by a spatially small 
quark-gluon wave packet ~\cite{CFS97}.

$\bullet$ 
Hard exclusive processes, like the production of vector 
mesons in DIS, are dominated by the interaction of $\gamma^*$ in small 
$q\bar q$ configurations ~\cite{Brodsky94}
Hence these processes provide a new
and effective method to measure the amplitude of small size 
quark-gluon configurations within wave functions of hadrons. 
Recently, the  minimal Fock component of the light- cone wave function of a 
pion has been measured at FNAL ~\cite{Ashery}
in the  process $\pi +A \to jet_1 + jet_2 +A$.

$\bullet$
~~
Different quark-gluon configurations within a fast 
hadron interact with a target with different strengths.
In particular, the cross section for  
colorless $q \bar q$ pair (the  triplet representation of 
$SU(3)_c$ where $F^2=4/3$ ) is different from a 
 colorless pair of gluons (the color octet 
representation of $SU(3)_c$ where $F^2=3$).

$\bullet$  
The Color Transparency Phenomenon in hard diffractive processes 
is a reasonable approximation to the solution of the QCD evolution equation  
in the limit of fixed $x$ but $b\rightarrow 0$ ~\cite{grenoble}:
\be
\sigma_{q\bar q A}(x,b)\rightarrow A 
\sigma_{q\bar q N}(x,b).
\label{CT}
\ee

$\bullet$ ~~
Within the $x, Q^2$ range of  applicability of the 
QCD evolution equation the 
cross section  of a small dipole interaction with a hadron 
target rapidly increases when $x\rightarrow 0$
and its  $A$ dependence becomes less steep as compared
to that within the CT  
regime where $\sigma_{q\bar q-A}=A\sigma_{q\bar q N}$.

~

{\it Data confirm this and several other striking QCD predictions in 
 hard exclusive  diffractive processes:}

~

$\bullet$  ~~
The fast increase  of cross sections of hard exclusive diffractive 
processes with energy follows from the  QCD factorization theorem 
\cite{CFS97}:
\be
\sigma_{hard~~ diffraction}\propto {\left(\alpha_s x 
G_{Target}(x,Q^2)\over Q^n\right)}^2, 
\label{FS89}
\ee
and $n\geq 2$. Such a  behavior has been predicted in the leading 
$\ln x$ approximation ~\cite{FS89,Ryskin93,FMS93,Blattel93,Brodsky94}. 
The  fast increase with energy has been observed in the diffractive 
electroproduction of $J/\psi$,$\phi$ and $\rho$ mesons~\cite{thisconference}.

$\bullet$ ~~ An onset of approximate flavor independence of cross 
sections at large $Q^2$  (for the same transverse color separation) 
as a consequence of flavor blindness of $QCD$ ~\cite{Brodsky94,Abramowicz94}  
has been observed in the ratio of $\phi$ and $\rho$ mesons yields
~\cite{thisconference}: $\sigma_(\gamma*+p\rightarrow \phi+p):
\sigma(\gamma*+p\rightarrow \rho+p)=2:9$, and in the increase  of the 
relative yield of $J/\psi$ and $\rho$ mesons with 
$Q^2$.~~\footnote{A larger value of $\left|\Psi(0)
\right|$ for mesons build of  heavier quarks leads to a prediction
of the high $Q^2$ enhancement of the $\phi/\rho, J/\psi/\rho$
ratios as compared to the $U(3)_f,U(4)_f$ expectations by the
factor of  1.2, 3.4,respectively.}

$\bullet$~~
The prediction of an almost universal and practically 
energy independent slope of the $t$ dependence of the 
differential cross section for any hard exclusive  
processes. This is because  a
hadron in such processes is predominantly produced in a configuration
much smaller than the radius of the hadron. The 
smallness of the derivative over $\ln 1/x$ of the slope
of $t$ dependence: $\alpha'(hard) \ll \alpha'(soft)$ 
is due to the hard physics occupying a  considerable part of
the  rapidity interval within a parton ladder ~\cite{Brodsky94}.
Both predictions have been 
recently confirmed by ZEUS ~\cite{Levy}.

$\bullet$~~
The generalized Color Transparency Phenomenon has been 
predicted for  the hard diffractive scattering of a 
$q \bar q$ pair of small transverse size $b$, 
off nuclei with a coherent nuclear recoil 
~\cite{FMS93,Brodsky94} as
\be
\frac{\sigma (q\bar{q}+A\rightarrow X+A)}{\sigma (q\bar{q}+N\rightarrow X+N)}
=\frac{r_{N}^{2}}{R_{A}^{2}}\frac{G_{A}(x,\frac{b_{o}^{2}}{b^{2}})}
{G_{N}(x,\frac{b_{o}^{2}}{b^{2}})}\rightarrow A^{4/3}\left|_{A\gg 1}\right..
 \label{CT1}
\ee 
This result should be valid at  fixed $x$ when $b^{2}\rightarrow 0$.

Here $r_N$, and  $R_A$ are radii of a nucleon (of a nucleus). 
The predicted $A$ dependence  is strikingly 
different from the A dependence of soft elastic (inelastic) 
coherent diffraction: $A^{2/3}$ ($A^{1/3}$). The expected 
strong dependence on atomic number in QCD  has been observed in 
$\gamma +A\rightarrow J/\psi+A$ -
$\propto A^{1.45}$  ~~\cite{sokoloff}. Very recently the $A$-dependence
$\propto A^{1.55\pm 0.05}$  for the process
$\pi +A\rightarrow 2jet +A$ was observed 
~\cite{Ashery} from the comparison of  platinum and carbon
targets. For this case eq.\ref{CT1} leads to $\sigma \propto A^{1.45}$. 
Note that the observed dijet production ratio for $Pt$ and 
$C$  is 8 times larger than the one measured in 
soft diffraction.  

~~~

~~~~
\subsection{Fast increase of Parton Densities contradicts to 
Probability Conservation.}
 
The key question for  future research is whether the cross section 
of the interaction of spatially small quark-gluon wave package is always 
small or whether it will become comparable with the $\pi N$ cross 
section of  at sufficiently small $x$ or even larger  as the 
interpolation of conventional parton distributions suggests.

To see that the growth predicted by eq.\ref{plc} cannot continue
forever, it is sufficient to consider the scattering of a trial hadron
representing a small dipole off a  hadron and to compare the total 
cross section of the inelastic scattering given by eq.\ref{plc} 
and  the  elastic cross section. The restriction follows from 
the inequality: $\sigma_{inel}(q\bar q T)\gg \sigma_{el}(q\bar q T)$
and the optical theorem: 
$\sigma_{el}(q\bar q T)=\frac{\sigma_{tot}(q\bar q T)^{2}}
{16 \pi B } (1+\eta^{2})$. Here  
$B$ is the slope of the elastic cross section
for the ``small dipole''-nucleon scattering. Experimentally  
$B\approx 4 GeV^{-2}$ and practically energy independent as follows 
from the measurement of hard diffractive processes ~\cite{thisconference}.
$\eta$ is the ratio
\begin{equation}
\eta= {Re A \over Im A}
\simeq \frac{\pi}{2}{d \over d\ln 1/x}  \ln \left(xG(x,Q^2)\right).
\label{reim}
\end{equation}
Another valuable restriction to the region of
applicability of DGLAP comes from the  requirement that 
cross sections of events with different multiplicities
should be positive. Taking into account  the screening due to
the double interaction of the small dipole with the target,
leads to 
\begin{equation}
\sigma_{inel}(q\bar qT)\geq 3\left[{\sigma_{el}(q\bar qT)+
\sigma_{diff}(q\bar qT)}\right].
\label{single}
\end{equation}  
Numerical calculations based on these inequalities 
and  conventional structure 
functions show that DGLAP should be certainly violated 
within the kinematics of pp  collision at LHC, and in $eA$ 
collisions already at HERA.

\subsection{Limiting behavior of parton densities at small x.}

With a decrease of $x$ the restriction by the LO and NLO terms in
$\alpha_s$ in the kernels of the DGLAP evolution equation  becomes 
suspicious because of the necessity to sum over all double logarithmic
terms $\ln(Q^2/Q_o^2)\ln x$. This is due to  the contribution of PQCD 
diagrams with gluon exchange being enhanced by powers of $\ln x$. 
This problem is not acute for the HERA kinematic
range where the permitted interval for the  PQCD evolution 
in rapidity is $\approx 7$ units for $x \sim 10^{-4}$ - therefore
the PQCD ladder  includes radiation of not more than 1-2 gluons.
The distance in rapidity between the adjacent partons in the ladder 
in the multiRegge kinematics is
at least 2-3 units and we exclude fragmentation regions of the 
projectile and the target $\approx \ln Q^2/Q_o^2+2~ \approx 4 $
units. For certainty we consider $Q^2=10 GeV^2$.
At the same time, at the LHC, where $x$ down to $10^{-7}$ 
can be studied, the  corresponding interval in rapidity can reach 14 units.
Thus  radiation of 3-4 gluons is possible in the kinematics
of LHC. So PQCD predicts that structure functions should be
polynomials over powers of $\ln x$. Maximal
power of $\ln x$ should be 2-3 at HERA and 4-5 at LHC.
But this PQCD gluon radiation leads to a  large parton density
which is in variance with probability conservation as has been 
explained above.

The expected physics and open questions can be visualized within the 
target rest frame description. A compact wave package is not an eigenstate 
of the QCD Lagrangian. Therefore it  evolves to a larger size
before hitting  the target as  a result of the Gribov diffusion in
the impact parameter space.  DGLAP and  BFKL approximations 
include such a diffusion. 
If  $\ln 1/x \gg \ln {Q^2/\Lambda_{QCD}^2}$   the number of steps in
the parton ladder in the longitudinal direction is much 
larger than in the  transverse one. Thus, as a result of Gribov 
diffusion in the transverse plane PQCD becomes inapplicable for 
the description of ultra-small $x$ behavior of the  parton distributions
since the system will expand to  a soft scale  long before hitting  
the target. 
(We want to draw attention that this reasoning does not imply 
violation of the QCD factorization  theorem for DIS in $ep$ collisions).
However small $x$ behavior can be hardly described by 
the Regge pole type formulae also
because the formation length of the Pomeron $\approx q_o/\mu^2$
is much larger than the distances important in DIS $\approx q_0/Q^2$.

One  option is  that the growth slows down when the  cross section
reaches a value much smaller than the black limit
$\sim \pi r_N^2$ (this value may decrease with the increase of $Q^2$)
and the further increase will follow the  pattern
of meson-nucleon interactions. In this case, PQCD physics will occupy 
a fraction of the whole phase volume which is independent of $x$.
The second option is
that the fast growth will continue all the way to the unitarity limit. In 
this case PQCD physics may occupy almost the whole phase volume. 
In both cases Bjorken scaling will be strongly violated.

The structure function of a target within the  black disc 
regime has been calculated by V.Gribov before QCD ~\cite{BH68}:  
\begin{eqnarray}
\sigma(\gamma*+T\rightarrow X)=
\frac{\alpha_{em}}{3\pi}
\pi r_{T}^{2} \int \rho(m^{2})m^{2} \frac{d m^{2}} 
{(m^{2}+Q^{2})^2}.
\label{blackdiff}
\end{eqnarray}
Here $\rho(m^{2})={\sigma(e\bar e\rightarrow hadrons)\over 
\sigma(e\bar e\rightarrow \mu \bar \mu)}$. 
Eq.\ref{blackdiff}~ leads to 
$\sigma(\gamma*+T\rightarrow X) \propto \ln (s/Q^2)$.
This is markedly different from the expectation of PQCD 
that the cross section should decrease as 
a 
power of $Q^{-2}$.

\section{Towards Superstrong Strong Interactions at the LHC}
\label{super}
\subsection{Coherence length and Interaction Strength 
 Fluctuations} 
It is well known that due to the uncertainty principle and the 
Lorentz slowing  of interaction between constituents within 
a fast projectile ``a'', the formation length of a  fast 
particle  ``a'' for a transition into 
state $|n>$ is given by the energy denominator:
\begin{equation}
l_{f}\approx \frac{1}{E_n-E_a}\approx 
\frac{2E_a}{m^{2}_{n}- m^{2}_{a}}\equiv
\frac{1}{2m_N x},
\label{form}
\end{equation}
where $E_a$ is the energy of particle $a$ in the rest frame of 
the target, and $m_{n}$ is the mass of state $|n>$.
For $x=10^{-3}$ (kinematics of HERA) $l_f=100 Fm$,
for $x=10^{-6}$ (kinematics of LHC) $l_{f}=10^{5} Fm$. 
Thus the wave function of fast projectile is build 
up at macroscopic (as compared to 1 Fm) 
distances. This  should lead to a variety of Coherence Phenomena. 

Perhaps the most straightforward application of eq.\ref{form} 
is the phenomenon of  fluctuations in the strength of hadron interactions, 
which among other things is 
responsible for the process of inelastic diffractive dissociation.
Indeed, it follows from the Lorentz slowing  of interactions 
within a fast projectile, h,  that high energy processes
should be described as the superposition of interactions of 
quark-gluon wave packages (instantaneous  quark-gluon configurations 
of the  projectile) with the weight given by the square of wave 
function of $h$. The crucial point is the  possibility to use the closure 
over the intermediate states, $\left| n\right>$. This is 
because the  minimal momentum transferred to target in inelastic diffraction
 $-t_{min}=(M_n^2-m_h^2)^2/s$
 tends to zero   for all transitions with  
$ M_n^2 \ll s$. Thus it is necessary to generalize the notion 
of a cross section to the distribution over cross 
sections - $P(\sigma)$, see  ~\cite{FMS} for review and references.
 Here the usual cross section is
average over strengths: $\sigma_{tot}\equiv 
<\sigma>= \int P(\sigma)\sigma d\sigma$.
The contribution into $P(\sigma)$ of small size configurations 
can be calculated by applying the QCD factorization theorem,
while the contribution of large (soft)
configurations has to be treated phenomenologically.
The dispersion over fluctuations of strengths - the  variance
of the  $P(\sigma)$ distribution, $\omega_{\sigma}$,  - 
can be inferred from the data using the formulae of ~\cite{MP78} for 
the ratio of inelastic and elastic cross sections in  forward scattering:
\be
\omega_{\sigma}=\frac{\frac{d\sigma(h+T\rightarrow X+T)}{dt}} 
{\frac{d\sigma(h+T\rightarrow h+T)}{dt}}
\left|_{t=0,X\neq h}\right.
=\frac{<\sigma^{2}>-<\sigma>^{2}}{<\sigma>^{2}}.
\label{Pumplin}
\ee
The analyses ~\cite{Blattel93,FMS93,FMS}
of diffractive processes of the proton and nuclei
at fixed target energies show (i) the important role of 
components with small $\sigma$ -  color transparency phenomenon;
(ii) the  significant probability of ``superstrong'' - larger than 
$\sigma_{tot}$ - interactions:
 $\int P_N(\sigma\geq \left<\sigma\right>)d\sigma \approx 50\%$,
$\int P_N(\sigma\geq 60 mb) d\sigma  \approx 20\%$.  
By itself, the very existence of such  superstrong strong
interactions is not surprising  since the large distance color forces 
are much more powerful than the usual internucleon forces. At the same time the
practical question arises what is the origin of superstrong 
interactions: strong meson fields, confinement forces between quarks at 
distances larger than average, etc and how does one
 investigate and  use them.

The generic property of QCD both in perturbative and nonperturbative
regimes is that the smaller the cross sections, the   faster is its  increase 
with energy (V.Gribov). As a result of the fast increase with energy of 
small cross sections  cf. eq. \ref{plc},  the distribution $P(\sigma)$
will become  narrower at larger energies. Therefore $\omega_{\sigma}$ should
decrease with energy. Indeed such a decrease is consistent with the
current data: $\omega_{\sigma}(\sqrt s=30 GeV)=0.3-0.35$,
$\omega_{\sigma}(\sqrt s=900 GeV)\sim 0.20$, and 
$\omega_{\sigma}(\sqrt s=1.8 TeV)\sim 0.15$. To reproduce this pattern
and the observed increase of $\sigma_{tot}(pp)$
starting from $P_N(\sigma)(\sqrt s=30 GeV)$, one also needs to assume 
that cross sections of interactions of superstrong configurations grow
with energy  $\propto \ln(s/s_0)$. (The impact parameter space framework
provides a related  interpretation of the decrease
of  $\omega_{\sigma}$ as due to the blackening of the interaction at 
the central impact parameters, see e.g. A.Mueller's talk
~\cite{Muellerdis98}.)
Overall, one expects a transition of high-energy strong
interactions  from multi-scale semihard/soft dynamics 
(which was due to the presence of color screening and the existence of
 small size configurations in hadrons)  to
a new soft dynamics with essentially one scale at  LHC energies.
This is close to the underlying picture of the Gribov Reggeon calculus,
though the Gribov picture assumed the validity of one scale 
dynamics at much smaller 
energies. However the  difference from the situation discussed 
at lower energies is that 
these new strong interactions are likely to be close to 
a phase transition, section \ref{softtrans}. 
Obviously, the one scale soft dynamics will be accompanied
by multiple hard interactions - productions of 2, 4, 6 ... jets. 

A related question is the rate of the energy variation  of 
the slope of the $t$ dependence of the elastic cross section
given by $\alpha'(s)$. At high energies small configurations
start to interact with large enough cross sections leading to
the onset of  Gribov diffusion for the interaction of these 
configurations as well. This physics would lead to an increase of  
$\alpha'$ with $s$. This may be  relevant
for  the difference of the value of  $\alpha' \sim 0.25 GeV^{-2}$ observed 
in $pp$ collisions at the 
Tevatron collider and $\alpha' \sim 0.15 GeV^{-2}$ observed
at HERA for  vector meson photoproduction and 
for  $\pi N$ scattering at fixed  target energies.
 $\alpha'$ should become universal at the energies where
all parton fluctuations ``forget'' about the state of the
  hadron which emitted them.

For a photon projectile the QCD factorization theorem predicts
 that  
$$P_\gamma(\sigma)\propto 1/\sigma$$
 for small $\sigma$.
With increase of $Q^2$ this term should dominate in 
$P_{\gamma^*}(\sigma)$.
On the contrary, hard and soft physics 
give comparable contributions to
$\sigma_{tot}(\gamma N)$. Soft physics is further enhanced in the diffractive
cross section.
Thus in order to understand the physics relevant for the 
fluctuations of strength of the interaction,   
it is necessary to concentrate efforts on the separate
investigation of phenomena where physics of small or
large size quark-gluon configurations would dominate.

\subsection{Soft QCD - on the way to a phase transitions?}
\label{softtrans}
Theoretical treatment of the physics relevant for   large $\sigma$
is still phenomenological although much of the gross features of soft QCD 
physics are well understood.  

Elastic and inelastic soft
diffractive
QCD processes are shadow of inelastic processes.  
The starting assumption of theoretical efforts is that 
hadron radiation in inelastic processes is a random process -no long range 
correlations between partons in rapidity space. This assumption is
in line with data at fixed target energies, leading 
to the hypothesis that high energy processes are dominated by the  
exchange of the Pomeranchuk trajectory ~\cite{Gribov61,chew-frautchi}.
Pomeranchuk trajectory  = parton  ladder where correlations 
between partons in rapidity are short range only~\cite{gribovdiffusion}.
Further theoretical investigation for the intercept of the Pomeron:
$\alpha(0)\geq 1$) found large long range 
correlations in rapidity between produced hadrons which are 
increasing with energy due to various 
multiPomeron type exchanges which could not be mimicked by an effective
single Pomeron exchange
~\cite{gribovreggeoncalculus}.

Experimental support of these theoretical ideas comes among other things from
 (i) the  observation at the S$p\bar p$S collider
of significant long range correlations in rapidity. 
(ii)   blackening of pp collisions at central impact parameters and 
(iii) the screening of  inelastic diffraction in the triple Pomeron
limit at collider energies.

The mathematical structure of the Gribov Reggeon calculus
for $\alpha(0)\geq 1$ resembles
the theory of  second order phase transitions near a critical point.
Indeed,  applying the V.Abramovski$\breve{{\rm i}}$, 
V.Gribov and O.Kancheli cutting rules ~\cite{AGK} one finds that
the analogous property would be very significant fluctuations 
of the density of the produced particles both for small and
large rapidity intervals.

It is also important that due to the Gribov diffusion
 the transverse size of the Pomeron ladder
reaches $\sim 1 fm$ at  LHC energies. Thus
multiPomeron  diagrams correspond to 
ladders with a strong spatial overlap.
The overlap effects  are further enhanced in 
$pA$ and in $AA$ collisions  at the LHC
since parton ladders from the interactions between different
nucleons strongly overlap in the center of rapidity
as a result of the Gribov diffusion in the impact parameter space
making the   formation of new forms of hadron matter
in AA collisions plausible.

\section{Experimental opportunities}
Above we outlined a number of outstanding problems in 
high-energy QCD which need to be addressed experimentally: physics 
of small $x$ i.e., large parton densities, longitudinal and transverse
multiparton correlations in hadrons, and  fluctuations of the interaction
strengths. In this section we will outline promising experimental
avenues for further investigations 
\footnote{For the  extensive discussion of  various experimental options in 
$pp$ and $pA$ collisions as well as relevant references see
FELIX LOI ~\cite{FELIX}.}.

\subsection{Where to search for the limiting behavior
of parton densities.}

It would be possible to push the studies of small x behavior of the
parton densities beyond the HERA range at higher energy $ep$ colliders,
for a summary see ~\cite{roeck}. Another opportunity would be a very
forward detector like FELIX at the LHC, which would be able to measure
quark densities down to $x \sim 10^{-7}$ and gluon densities down to 
$x \sim 10^{-6}$ ~~\cite{whitmore}.
Also, by measuring in the same detector hard scattering of two partons with 
$x_i \ge 0.2$ it would be possible to check whether QCD factorization
for hard scattering is still valid at the LHC, or the higher twist effects
due to high parton densities lead to effective screening
of hard parton interactions.

Two types of studies can be made at HERA.

$\bullet$ ~~ {\it Diffractive photon production in DIS at HERA}

It is possible to measure the ratio of the real and imaginary parts of
the scattering amplitude of the process $e+p\rightarrow e+\gamma +p$
via study of the azimuthal proton distribution of differential
cross sections.  
The relevant  term $\propto \cos(\phi(p)-\phi(e))$ arises
due to the interference between diffractive photon
production and the Bethe-Heitler process.
Numerical studies indicate that this effect is large enough to be
measured at HERA ~\cite{andreas}. Because of the relation between the
$Re/Im$ ratio and $d \ln xG(x,Q^2)/d \ln x$ - eq.\ref{reim},
this study would allow one  to  {\bf directly} measure
the rate of increase of gluon densities with energy.
This rate differs substantially for different models which fit 
$F_{2N}(x,Q^2)$ and hence such measurements would allow one  to
discriminate between possible models.

$\bullet$ The study of $eA$ scattering is feasible at HERA as well. 
The advantage of $eA$  collisions is that nonlinear effects 
in the QCD evolution equation are enhanced by the factor 
$A^{1/3}$ because $A^{1/3}$ nucleons  are at the same
impact parameter ~\cite{mq}. The enhancement would be even larger for
central impact parameters.

If nonperturbative shadowing of the
gluon distribution would be similar 
to that for $F_{2A}(x,Q^{2})$,   regime of large nonlinear QCD effects 
will be achieved in $eA$ collisions at HERA. The cross section
for black interaction has been calculated by Gribov, 
see eq ~\ref{blackdiff}.
 {\it Caution}: nonperturbative shadowing of gluon distributions
in nuclei is unknown at present.  If $\frac{G_{A}}{G_{N}}=A^{2/3}$,    
this may change the above conclusions. At the same time,  
it will change dramatically the theoretical expectations 
for hard  processes in $AA$ collisions at RHIC and LHC.

\subsection{Exclusive Hard Diffraction Possibilities}

We discussed above that the main question of small $x$
dynamics can be formulated as a question of dynamics of
the interaction of a small dipole with the target, hence the study of
 exclusive diffractive processes can provide the most direct
information on this issue.

\subsubsection{ep and eA processes}

In $ep$ scattering at  HERA and beyond,  this involves a more accurate 
study of exclusive vector meson production channels,
exclusive production  of two jets. The key question is the energy
dependence of the cross sections at fixed $Q^2$ or $p_t$ of the 
jet. Nonlinear effects would be manifested in a slowing down of the
energy dependence at the highest energies.
A crucial complementary
measurement would be a high precision study of the slope of the $t$
dependence. For large $Q^2$ the  slope should
be universal and independent of $Q^2$, and  the process. Thus 
$\alpha'_{eff} \ll \alpha'_{soft}$. The onset of soft dynamics
would lead to the emergence of the Gribov diffusion and hence
would  increase the slope.  

Since nonlinear effects are amplified in $eA$ collisions, the study
of the energy dependence of  coherent diffraction off nuclei at
HERA would be also very promising.  
Overall, these measurements together with the measurements of 
 nuclear structure  functions, provide one of the realistic methods
to observe the violation of the DGLAP approximation, and  investigate
the  onset the regime of large parton densities.

\subsubsection {Hard diffractive processes at LHC.}

To select scattering of a hadron in a small size configuration,
one needs a special trigger. It is provided by the  selection of the final
state where hadron is diffracting into a number of jets equal to the minimal
number of constituents in the hadron. The simplest reaction is 
$$\pi +T\rightarrow q(z,k_{t})+\bar q(1-z,-k_{t}) +T.$$ 
Here $z$ is the fraction of pion momentum carried 
by a quark and $k_{t}$ is its transverse momentum. The cross section of this 
process is calculable in terms of  the light-cone wave function of pion 
- $\phi(z)$~\cite{FMS}:
\be
\frac{d\sigma}{dt}={c|\phi(z)|^{2} 
\left(\alpha_{s} xG(x,k_{t}^{2})\right)^{2} \over k_{t}^{8}}
F^2_{2g}(t).
\label{piondiff}
\ee
The coefficient $c$ is calculable in QCD.
$F^2_{2g}(t) \approx \exp Bt, B \approx 4 \div 5 GeV^{-2}$ is the two gluon
form factor of the nucleon.
Experimentally~\cite{Ashery} for $k_{t}^{2}\geq 2GeV^{2}$
the pion wave function 
 is close to the asymptotic one:  $\phi(z,k_{t})\propto z(1-z)$
thus this reaction is an effective method to measure 
the  ``small dipole''-nucleon interaction cross sections.
At a collider, one can study this process using pion tagging:
$pp \to n(\Delta^{++}) + 2 jets + p$ ~\cite{FELIX}. 

Another, theoretically, clean option is the process
$p+p(A)\rightarrow jet(z_{1},k_{t1})
+jet(z_{2},k_{t2}) + jet(z_{3},k_{t3})+p(A)$. Its cross section  
is calculable in terms of wave function of the 
 minimal Fock $\left|3q \right>$
configuration within a proton if the relative $k_{ti}$
of the jets are large, as:
\begin{eqnarray}  
\frac{d\sigma(k_{t1}^{2})}{dz_{1} dz_{2} dz_{3} 
d^{2}k_{t1}  d^{2}k_{t2} d^{2}k_{t3}}=
 c_N |\alpha_{s} xG(x,Q^{2})|^{2}
\frac{|\phi_{N}(z_{1},z_{2},z_{3})|^{2}}{|k_{t1}|^{4}
|k_{t2}|^{4} |k_{t3}|^{4}}F^2_{2g}(t) \nonumber \\ 
\delta(\sum k_{ti}-\sqrt{-t}) 
\delta(\sum z_{i}).
\end{eqnarray}
The coefficient $c_N$ is also calculable in QCD. The numerical estimate
of the cross section integrated over all variables  except 
$p_t$ of one jet gives:
$$\sigma(3jets)\propto \frac{|\alpha_s xG(x,Q^2)|^{2}} 
{p_t^{8}} \propto 10^{-4\div -5} |\frac{5 GeV}{(p_{t})}|^{8}mb$$
for  LHC energies.  A pressing question is whether this cross section 
will grow up to the LHC energies as it is assumed in this estimate
based on extrapolations of $G_N(x,Q^2)$ to $x \sim 10^{-6}$.
Note that this process probes the minimal Fock three quark component
of the proton wave function, which is the same configuration 
relevant for the calculations of the proton decay.
A study of the same process with a  nuclear target would provide an
unambiguous test of the dominance of hard physics in this process.
Indeed for soft diffraction at the LHC, one expects the $A$ dependence
$\propto A^{0.25}$, while this process will grow at least as
$A^{0.7}$ assuming a maximum possible gluon shadowing of $G_A/G_N
\propto A^{2/3}$,  which is not likely.

\subsection{Fluctuations and correlation  phenomena}

\subsubsection{Long range rapidity correlations}
We discussed in  section 2 that $pp$ collisions may reach a 
situation close to a phase transition at LHC energies. 
Hence it would be necessary to study long range correlations in
rapidity for the  production of soft particles. Such correlations first 
emerge  due to double Pomeron exchange. The simplest  example of such
a correlation  is the correlation of backward and  forward
multiplicities observed at the S$p\bar p$S collider. the large rapidity
interval available at the  LHC would allow
one  to study these and higher order 
correlations using a  large acceptance detector.

It is important to perform similar studies in deep inelastic
scattering to search for effects of the exchange of  multiple soft ladders.
Current HERA detectors have rather limited acceptance in $y$ so the 
only practical approach seems to be a study of the correlation between the
yield  of the leading baryons at $x_F \sim 0.5-0.8$ and the central
hadron multiplicity for multiplicities
much larger than average ~\cite{FSneutron}. If the forward coverage is
improved many other options would become feasible.

\subsubsection{Geometry of Multijet production}
\label{multi}
At high energies the  cross section of multiple parton collisions becomes
large, so it is feasible to measure double(triple) parton distributions in 
the process of production of four(six)  jets in the
kinematics where $p^{jet,forward,1}_t \sim -p^{jet, backward,1}_t$,
 $p^{jet,forward,2}_t \sim -p^{jet, backward,2}_t$.
The probability of these processes is sensitive to the longitudinal and
transverse correlations between the partons with a 
resolution $\sim p^{jet}_t$. Crucial for such studies is a 
sufficiently large rapidity interval to separate these 
processes from the processes of two parton collisions with a production
of several jets.  Recently  CDF  has reported ~\cite{double}
a  cross section for double collisions which exceeds
the   expectations based on the model where partons are distributed
randomly in the impact area of the nucleon by a factor of
$\sim 2$. It  appears that the 
study of such cross sections may be one of the important steps in 
determining the three-dimensional structure of the nucleon. 

Also, the selection of events with multiple collisions strongly reduces the
average impact parameter for a collision. Hence, high density phenomena
should be manifested  much more prominently in  multiparton
collision events and can be used to  experimentally  distinguish
peripheral and central $pp$ collisions.

\subsubsection{Breakdown of Pomeron type factorization in inclusive 
diffraction}

One of the distinctive features of diffractive physics is the interplay 
between soft and hard QCD. Thus the  investigation of diffractive processes 
in DIS may help to elucidate physics relevant to the origin of 
the Pomeron. A key property of the Pomeron exchange is the Regge pole 
factorization- lack of communication between projectile and target 
fragmentation regions. On the contrary,  QCD factorization is valid for 
the interaction  of spatially small configurations only. The mismatch 
between two factorizations is the source of many new phenomena. 

In the case of inclusive diffraction in DIS one can express the 
cross section of diffractive scattering at fixed $x_p$  (Feynman $x$ 
of the proton) in the leading twist through the  $Q^2$ dependent 
parton densities, $f_{diff}^i(\beta,Q^2,x_p)$ which are often referred 
to as  Pomeron (anti)quark and gluon densities. However in contrast with
 the usual parton densities which are process independent, a 
``Pomeron parton density'' is not a universal object since the soft 
screening interactions depend both on the target and on the projectile. 
At HERA this phenomenon can be studied in resolved photon
hard diffraction where screening effects should reduce
$f_{diff}^i$ as compared to the case of the DIS diffraction.
Striking nonuniversality should be present due to the filtering
phenomenon and presence of  the $\gamma^*$ wave function with the masses
$\sim Q^2$ in $eA$ diffraction ~\cite{FSAGK}. In particular,
$f_{diff}^i(\beta,Q^2)$ should strongly 
depend on the atomic number of the target increasing at $\beta \ge
0.4$ and decreasing at small $\beta \le 0.1$.

In $pp$ collisions the  strong suppression of $f_{diff}^i$ has been 
already observed in  experiments at the 
FNAL collider both in two jet, and $W-$boson production channels.
Here the challenging problem is to try to observe the breakdown of
factorization as  a function of the $x_1$ fraction carried
by the hard parton  of the diffracting proton. 
Such a  dependence seems natural due to fluctuations of the strength of the 
interactions which we discussed in section \ref{super}. Indeed if for large
$x_1$, configurations with smaller $\sigma$ are selected,  screening
effects are reduced and $f_{diff}^i$, extracted from the data, would be larger.

\subsection{New phenomena in pA collisions at the LHC}

It is feasible to study collisions of protons and nuclei at 
the LHC. Such
studies would be of utmost importance for the interpretation of
the LHC heavy-ion collision data as well the cosmic rays of ultra-high energies.
The transverse area  of the proton at these energies is $\sim$  4
times larger than at fixed target energies, and thus,  hard processes
 are dramatically enhanced. Therefore  one expects a very
different picture of $pA$ interactions at these energies from the one
explored up to now at $\sqrt(s) \le 40 GeV$.
It would be possible to investigate  many qualitatively 
new phenomena. We list here
several of them.

$\bullet$~~ {\it Nonlinear phenomena} via the study of nuclear shadowing at
$x$ down to $10^{-7}$.

$\bullet$~~ {\it Parton propagation in nuclear media} via
the  study of $p_t$ - broadening of leading  Drell-Yan pairs,
dijets in the proton fragmentation region, and parton energy losses.
For the summary of  QCD expectations see ~\cite{yura}.

$\bullet$ ~~{\it Multiparton distributions in a  proton} via 
the study of double (triple) hard parton scattering. Comparison of 
the yields for $pp$ and $pA$ collisions would allow one 
to measure the transverse separation
of hard partons  in a model independent way.

$\bullet$ ~~{\it Propagation of small color dipoles through nuclear
  media} would allow one to address the question whether
 interactions of small objects may become comparable to the
interaction of ordinary hadrons  at super-high
energies. Tools to investigate this question
include (i) hard exclusive diffraction of proton into three jets
which  we discussed above, and (ii) the production of 
{\it a leading} $J/\Psi$. In the later case, one may expect a 
strong change of  the $A$ dependence of the 
leading $J/\psi$ meson 
production in $pA$ collisions as compared to the 
fixed target energies where 
$\sigma(pA\rightarrow J/\psi+X)\propto A$
if $x_F(J/\psi) \ll 1$.
Two effects are relevant: a  large gluon shadowing for small 
$x \sim M_{J/\Psi}^2/s$, and the large cross
section for the interaction of a $c\bar c$ pair of the size $\sim 0.2
fm$ with the nuclear target.
Overall this may lead to a reduction of the $A$-dependence to as low as
$A^{1/3}$. A less dramatic but still significant effect may be expected
for $ \Upsilon$ production.

$\bullet$ {\it Color fluctuations in protons} via the study of 
correlation  properties between  
soft hadron production and the presence of hard collision at large $x_p$.
The expectation is that for $x_p \ge 0.5$, configurations in the colliding
proton
with a large interaction strength $\sigma \ge \sigma_{tot}$ are strongly
suppressed, leading to a significant reduction of soft hadron
multiplicity, etc ~\cite{FS85}.

\section*{Acknowledgments}

We wish to express our sincere gratitude to
J.Bartels, J.Bjorken, V.Gribov, A.Mueller, for the fruitful 
discussions of pressing problems of QCD, to V.Guzey for the 
calculation of the boundary of applicability of the QCD evolution equation 
to $eA$ scattering. This work was supported by the U.S. Department of 
Energy under Contract No. DE-FG02-93ER40771, 
by the Israel Academy of Science under contract N 19-971.

\end{document}